\documentclass[12pt,a4paper]{iopart}
\usepackage{iopams}
\usepackage{epsfig}
\usepackage{url}
\usepackage[dvips]{color}
\newcommand{\eqref}[1]{(\ref{#1})}

\definecolor{Black}{rgb}{.0,.0,.0}
\definecolor{Red}{rgb}{1.,0.,0.}

\begin{document}

\title{The radial BAO scale and Cosmic Shear, a new observable for Inhomogeneous Cosmologies}
\author{Juan Garc\'{\i}a-Bellido$^1$, Troels Haugb{\o}lle$^{2,1}$}
\address{$^1$ Instituto de F\'{\i}sica Te\'{o}rica UAM-CSIC,
Universidad Aut\'{o}noma de Madrid, Cantoblanco, 28049 Madrid, Spain,\\
$^2$ Department of Physics and Astronomy, University of Aarhus, DK-8000
Aarhus C, Denmark}
\ead{juan.garciabellido@uam.es, haugboel@phys.au.dk}

\begin{abstract}
As an alternative explanation of the dimming of distant supernovae it
has recently been advocated that we live in a special place in the
Universe near the centre of a large spherical void described by a
Lema\^itre-Tolman-Bondi (LTB) metric. In this scenario, the Universe is 
no longer homogeneous and isotropic, and the apparent late time 
acceleration is actually a consequence of spatial gradients. We propose 
in this paper a new observable, the normalized cosmic shear, written in 
terms of directly observable quantities, and calculable in arbitrary 
inhomogeneous cosmologies. This will allow future 
surveys to determine whether we live in a homogeneous universe or not.
In this paper we also update our previous observational constraints 
from geometrical measures of the background cosmology. We include the 
Union Supernovae data set of 307 Type Ia supernovae, the CMB acoustic
scale and the first measurement of the radial baryon acoustic oscillation
scale. Even though the new data sets are significantly more constraining, 
LTB models -- albeit with slightly larger voids -- are still in excellent 
agreement with  observations, at $\chi^2/{\rm d.o.f.} = 307.7/(310-4)=1.005$.
Together with the paper we also publish the updated {\tt easyLTB} 
code~\footnote{The code can be downloaded at
\url{http://www.phys.au.dk/~haugboel/software.shtml}} used for 
calculating the models and for comparing them to the observations.
\\ $\phantom{ }$\hspace{\stretch{1} Preprint: IFT-UAM/CSIC-08-69}
\end{abstract}

\submitto{JCAP}
\maketitle

\section{Introduction}

The nature of the matter and energy contents of the Universe is still a mystery.
While many cosmological parameters are known to a few percents precision --
which suggests we can start to speak of a Standard Cosmological Model --
some of these parameters correspond to quantities for which we completely 
ignore its nature. According to this model, only 4\% of the Universe is made of
something whose nature we know. What is the rest made off? Moreover, our 
current model of the Universe is based on a set of symmetries we have never 
tested very thoroughly, like homogeneity and isotropy, and some assumptions
we have taken for granted, perhaps too lightly, like the generalized Copernican
Principle, also known as the Cosmological Principle. These symmetries and 
assumptions provide the theoretical basis for the nowadays predominant
homogeneous and isotropic Friedmann-Robertson Walker (FRW) model of
the Universe.

However, distant supernovae appear dimmer than expected in a matter-dominated
FRW Universe.  The currently favored explanation of this dimming is the late 
time acceleration of the Universe due to a mysterious energy component that 
acts like a repulsive force. Observations seem to suggest that this so-called 
Dark Energy is similar to Einstein's cosmological constant, 
but there is inconclusive evidence. There has been a
tremendous effort in the last few years to try to pin down deviations
from a cosmological constant using deep galaxy catalogues like 
2dFGRS~\cite{2dFGRS} and SDSS~\cite{SDSS}, and extensive
supernovae surveys like ESSENCE~\cite{ESSENCE}, SNLS~\cite{SNLS}, 
UNION~\cite{UNION}
and SDSS-SN~\cite{SDSS-SN}, and many more are planned for the near 
future e.g.~DES~\cite{DES}, PAU~\cite{PAU,Benitez2008}, BOSS~\cite{BOSS}
and LSST~\cite{LSST}.

In the meantime, the realization that the universe around us is actually far
from homogeneous and isotropic has triggered the study of alternatives to this 
mysterious dark energy.  Since the end of the nineties it has been suggested by 
various groups~\cite{Hellaby:1998,Celerier:1999hp,Tomita:2000jj,Moffat:2005yx,
Alnes:2005rw,Garfinkle:2006sb,Enqvist:2006cg,Enqvist:2007vb,Mattsson:2007tj,
Wiltshire:2007zj,Alexander:2007xx,GBH:2008a,GBH:2008b,Zibin:2008b,Clifton:2008hv}
that an isotropic but inhomogeneous Lema\^itre-Tolman-Bondi (LTB)
universe could also induce an apparent dimming of the light of distant
supernovae, in this case due to local spatial gradients in the
expansion rate and matter density, rather than due to late time
acceleration. It is just a matter of interpretation which mechanism is
responsible for the dimming of the light we receive from those
supernovae. Certainly the homogeneous and isotropic FRW model is more
appealing from a philosophical point of view (it satisfies the Cosmological
Principle, i.e. it assumes spatial sections are maximally symmetric), but so 
was the static universe and we had to abandon it when the recession of 
galaxies was discovered at the beginning of last century.

There is nothing wrong or inconsistent with the possibility that we
live close to the centre of a gigaparsec-scale void. Such a void may
indeed have been observed as the CMB cold
spot~\cite{Cruz:2006sv,Cruz:2006fy,Cruz:2008sb} and smaller voids
have been seen in the local galaxy distribution \cite{Frith:2003,Granett:2008}.
The size and depth of the {\em distant} voids, i.e.~$r_0 \sim 2$ Gpc
and $\Omega_M \sim 0.2$, within a flat Einstein-de Sitter universe,
seems to be consistent with that in which we may happen to
live~\cite{Tully:2007tp}, and such a local void could account for the supernovae
dimming, together with the observed Baryon Acoustic Oscillations (BAO),
the CMB acoustic scale, the age of the universe, the local rate of expansion,
etc.~\cite{GBH:2008a}.

Observations suggest that if there is such a large void, we should
live close to the centre, otherwise our anisotropic position in the void would 
be seen as a large dipole in the CMB \cite{Alnes:2006pf}.  Of course, we do observe
a dipole, but it is normally assumed to be due to the combined gravitational pull of
the large scale structure in the local universe, such as the Virgo cluster, and the
Shapley super cluster. There is always the
possibility that we live off-centre and we are moving towards the
centre of the void, so that the two effects are partially cancelled,
giving rise to the observed dipole. However, such a coincidence could
not happen for all galaxies in the void and, in general, clusters that
are off-centred should see, in their frame of reference, a large CMB
dipole. Such a dipole would manifest itself observationally for us as an 
apparent kinematic Sunyaev-Zeldovich (kSZ) effect for the given cluster.
The still preliminary observations of the kSZ effect allowed us 
recently~\cite{GBH:2008b}
to put some bounds on LTB models. It is expected that
in the near future, improved measurements of this effect with the Atacama
Cosmology Telescope (ACT)~\cite{ACT}  and the South Pole Telescope
(SPT)~\cite{SPT} will very strongly constrain the LTB model. 

In fact, what we require are new data sets that could be used to constrain
further the LTB models and eventually rule them out, if possible. We thus
use the recently released compilation of 307 supernovae by the Supernova 
Cosmology Project - UNION collaboration~\cite{UNION}. Moreover, in this
paper we make use of a new and very interesting observation of the
baryon acoustic oscillation along the line of sight, otherwise known as
radial BAO (RBAO)~\cite{RBAO,GMS}. It provides a direct measurement of
the rate of expansion along the line of sight, $H_L(z)$, whose integral 
determines the luminosity distance and could in principle be different 
from that transverse to the line of sight, $H_T(z)$, which relates to 
the angular diameter distance. Both appear in the Einstein equations and
it is possible to disentangle their respective effects by looking at
a new observable, the ratio of shear to expansion in LTB models.
In FRW universes, the shear of the background geometry is identically
zero, while in LTB models it can be significantly different from zero, and 
becomes maximal at a certain redshift, corresponding to the size of the void. 
Therefore, by measuring the normalized shear, future galaxy surveys will be 
able not only to determine whether we live in a homogeneous universe or not,
but to measure the size and depth of the void.

In section~2 we describe the general LTB void models, giving the corresponding
Einstein-Friedmann equations, as well as parameterizations of their solutions. 
In a subsection we describe the GBH constrained model, where we
assume the Big Bang is homogenous, and thus the model only depends on a single
function, the inhomogeneous matter ratio $\Omega_M(r)$. In section~3 we
introduce a new observable, the normalized cosmic shear, 
which could help distinguish observationally
homogeneous FRW models from arbitrary inhomogeneous universes.
In section~4 we show how to properly calculate the size of standard rulers in
LTB models, constructing an effective Einstein de Sitter background, that can
be used for calculating early universe
quantities in the corresponding LTB model using standard formulae. In 
section~5 we analyse present observations of BAO along the line of sight, as
well as the recent UNION compilation of Supernovae, and give constraints on the 
model from current observations. Finally, in section~6 we give a discussion of
future prospects and some conclusions.

\section{Lema\^itre-Tolman-Bondi void models}
The LTB model describes general radially symmetric
space-times and can be used as a toy model for describing voids in the universe
\cite{Lemaitre:1997ab,Tolman:1934za,Bondi:1947av}.
The metric is
\begin{equation}\label{eq:metric}
ds^2 = - dt^2 + X^2(r,t)\,dr^2 + A^2(r,t)\,d\Omega^2\,,
\end{equation}
where $d\Omega^2 = d\theta^2 + \sin^2\theta d\phi^2$.
Assuming a spherically symmetric matter source with negligible pressure,
\begin{equation}
T^\mu_{\ \nu} = - \rho_M(r,t)\,\delta^\mu_0\,\delta^0_\nu\,,
\end{equation}
the $(0,r)$ component of the Einstein equations, $G^0_{\ r} = 0$, implies
$X(r,t)=A'(r,t)/\sqrt{1-k(r)}$, with an arbitrary function $k(r)$
playing the role of the spatial curvature parameter.
The other components of the Einstein equations read \cite{Enqvist:2006cg,
Enqvist:2007vb,GBH:2008a}
\begin{eqnarray} \label{eq:FRW1}
H_T^2 + 2H_TH_L + {k\over A^2} + {k'(r)\over A A'} = 8\pi\,G\,\rho_M \,, \\
2{\dot H}_T + 3H_T^2 + {k \over A^2} = 0\,,
\end{eqnarray}
where $\dot{\phantom{i}} \equiv \partial_t$ \& $' \equiv \partial_r$, and we
have defined the transverse and longitudinal Hubble rates as
$H_T\equiv \dot A / A$, and $H_L\equiv \dot A' / A'$.
Integrating the last equation, we get
\begin{equation}
H_T^2 = {F(r)\over A^3} - {k(r)\over A^2}\,,
\end{equation}
with another arbitrary function $F(r)$, playing the role of effective
matter content, which substituted into the first equation gives
\begin{equation}
{F'(r)\over A'A^2(r,t)} = 8\pi\,G\,\rho_M(r,t)\,,
\end{equation}
where $\rho_M(r,t)$ is the physical matter density.

The boundary condition functions $F(r)$ and $k(r)$ are specified by
the nature of the inhomogeneities through the local Hubble rate, the
local total energy density and the local spatial curvature,
\begin{eqnarray}
F(r) = H_0^2(r)\,\Omega_M(r)\,A_0^3(r)\,, \\
k(r) = H_0^2(r)\Big(\Omega_M(r)-1\Big)\,A_0^2(r) \,,
\end{eqnarray}
where functions with subscripts $0$ correspond to present day values,
$A_0(r) = A(r,t_0)$ and $H_0(r) = H_T(r,t_0)$. With these definitions, the
$r$-dependent transversal Hubble rate can be written 
as~\cite{Enqvist:2006cg,Enqvist:2007vb}
\begin{equation}\label{eq:hubblerate}
H_T^2(r,t) = H_0^2(r)\left[\Omega_M(r)\left({A_0(r)\over A(r,t)}\right)^3 +
(1-\Omega_M(r))\left({A_0(r)\over A(r,t)}\right)^2\right]\,,
\end{equation}
and we fix the gauge by setting $A_0(r)=r$. For fixed $r$ the above equation
is equivalent to the Friedmann equation of an open universe, and have an exact
parametric solution, while also very good approximate solutions can be found by
Taylor expanding around an Einstein de Sitter solution (see \cite{GBH:2008a} for details).

For light travelling along radial null geodesics, $ds^2=d\Omega^2=0$, we have
\begin{equation}
{dt\over dr} = \mp {A'(r,t)\over\sqrt{1-k(r)}}
\end{equation}
which, together with the redshift equation,
\begin{equation}\label{eq:null}
{d\log(1+z)\over dr} = \pm {\dot A'(r,t)\over\sqrt{1-k(r)}}
\end{equation}
can be written as a parametric set of differential equations,
with $N=\log(1+z)$ being the effective number of e-folds
before the present time,
\begin{eqnarray}\label{eq:lightrayt}
&&{dt\over dN} = - {A'(r,t)\over\dot A'(r,t)} \,,\\ \label{eq:lightrayr}
&&{dr\over dN} = \pm {\sqrt{1-k(r)}\over\dot A'(r,t)}
\end{eqnarray}
Notice, that while the angular diameter distance $d_A = A(r,t)$
only depends on the integral of the transversal Hubble rate
$H_T(r,t)$, the comoving and the luminosity distances 
$d_L = (1+z)\,d_C = (1+z)^2 d_A$ depend through
Eq.~\eqref{eq:lightrayt} on a mixture of $H_T$ and $H_L$.

\subsection{The constrained GBH model}
In general LTB models are uniquely specified by the two functions
$k(r)$ and $F(r)$ or equivalently by $H_0(r)$ and $\Omega_M(r)$, but
to test them against data we have to parameterise the functions, to
reduce the degrees of freedom to a discrete set of parameters.  For
simplicity in this paper we will use the constrained GBH model
\cite{GBH:2008a} to describe the void profile. First of all, it uses a
minimum set of parameters to make a realistic void profile, and
secondly, by construction, the time to the Big Bang is constant for
spatial slices, as would be expected in a generic model of inflation. 
Moreover, void models with an inhomogeneous Big Bang contain a mixture
of growing and decaying modes, and consequently the void does not 
disappear at high redshift making them incompatible with the
Standard Big Bang scenario~\cite{Zibin:2008a}.
If we only consider constrained LTB models, then at
high redshifts and/or at large distances the central void is reduced
to an insignificant perturbation in an otherwise homogeneous universe
described by an FRW metric, and physical results for the early universe
derived for FRW space-times still hold, even though we are considering
an LTB space-time.

The second condition gives a relation between $H_0(r)$
and $\Omega_M(r)$, and hence constrain the models to one free
function, and a proportionality constant describing the overall
expansion rate. Our chosen model is thus given by
\begin{eqnarray}
\hspace{-2cm}&&\Omega_M(r) = \Omega_{\rm out} + \Big(\Omega_{\rm in} - 
\Omega_{\rm out}\Big)
\left({1 - \tanh[(r - r_0)/2\Delta r]\over1 + \tanh[r_0/2\Delta r]}\right) \\
\hspace{-2cm}&&H_0(r) = H_0\left[{1\over \Omega_K(r)} -
{\Omega_M(r)\over\sqrt{\Omega_K^3(r)}}\ {\rm sinh}^{-1}
\sqrt{\Omega_K(r)\over\Omega_M(r)}\right] =
H_0 \sum_{n=0}^\infty {2[\Omega_K(r)]^n\over(2n+1)(2n+3)}\,,
\end{eqnarray}
where $\Omega_K(r) = 1 - \Omega_M(r)$, and the second equation follows
from the requirement of a constant time to a homogeneous Big Bang. We
use an ``inflationary prior'', and assume that space is asymptotically
flat, i.e.~in the following we set $\Omega_{\rm out}=1$. The model has
then only four free parameters: The overall expansion rate $H_0$, the
underdensity at the centre of the void $\Omega_{\rm in}$, the size of
the void $r_0$, and the transition width of the void profile $\Delta
r$. For more details on the model see Ref.~\cite{GBH:2008a}.

\section{The Raychaudhuri equation and the normalized shear}\label{sec:sigma}

Now that cosmological surveys are beginning to provide detailed maps
of the universe up to significant distances we can start to consider
measuring deviations from homogeneous FRW models. The main difference 
between FRW and LTB models is that the latter introduces two different
components to the rate of expansion: There is both a transverse 
(perpendicular to the line of sight) and longitudinal (along the line of sight) 
components. This induces a differential growth of the local volume of the universe.
Let us see how to quantify this geometrical distortion in the expansion of the universe.

Suppose we start with a set of comoving observers localized on a
sphere of volume $V$ in a LTB universe. Each observer has a unit vector
$n^\mu$ tangent to its trajectory, and follows a geodesic of the metric,
i.e. $n^\mu n^\nu_{\ ;\mu} = 0$. If we now consider a congruence of 
observers (i.e. geodesics) we can follow the change in positions of 
the observers with time. The covariant derivative of the congruence 
can be split into three distinct components,
\begin{equation}\label{theta}
\Theta_{\mu\nu} = n_{\mu;\nu} = {1\over3}\Theta P_{\mu\nu} + 
\sigma_{\mu\nu} + \omega_{\mu\nu}\,,
\end{equation}
where $P^\mu_{\ \nu} = \delta^\mu_{\ \nu} + n^\mu n_\nu$ projects out 
any tensor into a plane orthogonal to the congruence $n^\mu$. The tensor
$\Theta_{\mu\nu}$ measures the extent to which neighbouring trajectories
deviate from remaining parallel. The three components of $\Theta_{\mu\nu}$
in Eq.~(\ref{theta}) have different physical meanings. The trace gives 
the expansion rate of the congruence, $\Theta= n^\mu_{\ ;\mu}$, and
characterizes the growth of the overall volume $V$ of the sphere of
observers. The traceless symmetric part is the $shear$ tensor of the 
congruence, 
\begin{equation}\label{shear}
\sigma_{\mu\nu} = {1\over2}P^\alpha_{\ \mu}P^\beta_{\ \nu}
(n_{\alpha;\beta}+n_{\beta;\alpha}) - {1\over3}\Theta P_{\mu\nu}\,,
\end{equation}
and represents the distorsion in the shape of the sphere of test
particles (observers), giving rise to an ellipsoid with possibly
different axes' lengths but same volume $V$. Finally, the antisymmetric
part is the $vorticity$ tensor,
\begin{equation}\label{vorticity}
\omega_{\mu\nu} = {1\over2}P^\alpha_{\ \mu}P^\beta_{\ \nu}
(n_{\alpha;\beta}-n_{\beta;\alpha})\,,
\end{equation}
which describes the rotation of the sphere of test particles around
the center, again without changing the volume of the sphere.
Note that these are purely geometrical quantities and are independent 
of the dynamics responsible for time evolution in this metric.
By computing the covariant derivative of $\Theta_{\mu\nu}$ along the
worldlines of observers and taking the trace, one finds the famous
Raychaudhuri equation,
\begin{equation}\label{Raychaudhuri}
{d\Theta\over d\tau} = -{1\over3} \Theta^2 - \sigma_{\mu\nu}\sigma^{\mu\nu} 
+ \omega_{\mu\nu}\omega^{\mu\nu} - R_{\mu\nu}n^\mu n^\nu\,.
\end{equation}
The last term of the equation is the only one sensitive to the theory
of gravity. For general relativity and geodesic observers it becomes
$R_{\mu\nu}n^\mu n^\nu = 8\pi G(T_{\mu\nu} - {1\over2}g_{\mu\nu}T)n^\mu n^\nu = 
4\pi G(\rho + 3p)$. It is this term which allows for accelerated expansion 
in a $\Lambda$CDM-FRW model (for which both shear and vorticity vanish).

We can now compute all the terms in this equation in our LTB models. The 
global rate of expansion becomes $\Theta = H_L + 2H_T$; the spatial shear 
is non-vanishing, $\sigma_{ij} = (H_T - H_L)\,M_{ij}$, where $M^i_j=
{\rm diag}(-2/3,1/3,1/3)$ is a traceless symmetric matrix; while the vorticity
tensor vanishes, $\omega_{ij} = 0$. Note that the shear component of the congruence
vanishes asymptotically as we approach the Einstein-de Sitter universe, as
well as locally, close to the center of the LTB void. However at intermediate
redshifts it has its maximal deviation from zero, where the longitudinal
and transverse rates of expansion differ maximally, corresponding to the
edge of the void. This shear component is
a distinctive feature of inhomogeneous cosmological models, in fact it is one 
that could be used to distinguish FRW from LTB models in the near future with the
next generation of astronomical surveys like DES, PAU, BOSS and LSST.

In order to be quantitative, we should find a directly measurable quantity
that can be computed and compared with observations. From a theoretical
point of view, the natural variable to consider is the ratio of shear to expansion,
\begin{equation}
\varepsilon \equiv \sqrt{3\over2}{\sigma\over\Theta} = {H_T-H_L\over H_L + 2H_T}\,,
\end{equation}
where $\sigma^2 \equiv \sigma_{ij}\sigma^{ij} = {2\over3}(H_T-H_L)^2$. 
However, these rates are difficult to measure directly (in particular the
transverse rate of expansion $H_T$ can only be measured indirectly). More 
appropriate would be to use physical quantities like angular diameter 
distances and radial BAO scales, as a function of redshift or its derivatives. 

One can indeed find a relation between the
variable $\varepsilon$ and those quantities, by noting that in LTB models,
the angular diameter distance $d_A(z)$ is nothing but $A(r,t)$ in the 
metric ~(\ref{eq:metric}). Its derivatives give the various rates of expansion,
$H_T = \dot A/A$ and $H_L=\dot A'/A'$, but note that in general, the time
derivatives in an inhomogeneous universe are not directly related to the
redshift derivatives since there is also a contribution from spatial gradients,
$${d\over dN} = {\sqrt{1-k(r)}\over A'H_L}{d\over dr} - 
{1\over H_L}{d\over dt}\,,$$
where we have used Eqs.~(\ref{eq:lightrayt}-\ref{eq:lightrayr}), and therefore
$H_L$ and $H_T$ are not directly measurable quantities, contrary to the
case in FRW models.  Computing explicitly the derivatives of $A(r,t)$ with
respect to redshift, we find 
\begin{equation}
{H_T\over H_L} = {\sqrt{1-k(r)}\over AH_L} - {d\ln A\over dN} =
{\sqrt{1-k(r)}\over d_A(z)H_L(z)} - {d\ln d_A(z)\over d\ln(1+z)}\,,
\end{equation}
which gives for the new variable, the normalized shear $\varepsilon$, as a
function of the line of sight rate of expansion and the angular diameter distance,
\begin{equation}\label{nshear}
\varepsilon = {\sqrt{1-k(r)} - H_L(z)\,\partial_z[(1+z)\,d_A(z)]
\over 3H_L(z)d_A(z) + 2\sqrt{1-k(r)} - 2H_L(z)\,\partial_z[(1+z)\,d_A(z)]}\,.
\end{equation}
This quantity still depends on an unknown, the spatial curvature $k(r)$. However, 
we can integrate Eq.~(\ref{eq:lightrayr}) to give
$$\sqrt{1-k(r)} = \cosh\Big[H_0(r)\sqrt{\Omega_K(r)}\,\int_0^z {dz'\over
H_L(z')}\Big]\approx 1 + {1\over2} H_0^2(r)\Omega_K(r)\Big(\int_0^z
{dz'\over H_L(z')}\Big)^2 \,,$$
which is not far from unity in a wide range of LTB models that agree with
observations, see Ref.~\cite{GBH:2008a,GBH:2008b}.
In this case, the normalized shear becomes
\begin{equation}\label{nshearapp}
\varepsilon(z) \approx {1 - H_L(z)\,\partial_z[(1+z)\,d_A(z)]
\over 3H_L(z)d_A(z) + 2 - 2H_L(z)\,\partial_z[(1+z)\,d_A(z)]}\,,
\end{equation}
which can in principle be measured by any of the future surveys. In particular,
the rate of expansion along the line of sight has recently been measured 
by Gazta\~naga et al. at two different redshifts using the radial BAO 
scale~\cite{RBAO,GMS}, and together with measurements of angular diameter
distances could give a value for $\varepsilon(z)$. Note that in FRW universes
the normalized shear~(\ref{nshear}) vanishes identically since $H_L=H_T=H$ and 
$$(1+z)d_A(z)={1\over H_0\sqrt{\Omega_K}}\sinh\Big[H_0\sqrt{\Omega_K}
\int_0^z {dz'\over H(z')}\Big]\,.$$ Note also that the function $H_L(z)d_A(z)$ in the 
denominator is nothing but the Alcock-Paczynski factor $f(z)$, which is used as a
geometric test for the existence of vacuum energy in $\Lambda$CDM FRW models.
On the other hand, it is worth making emphasis on the fact that this normalized
shear~(\ref{nshearapp}) is independent of the value of $H_0$, still a mayor 
uncertainty in cosmology.

\begin{figure}[t]
\begin{center}
\includegraphics[width=0.65 \textwidth]{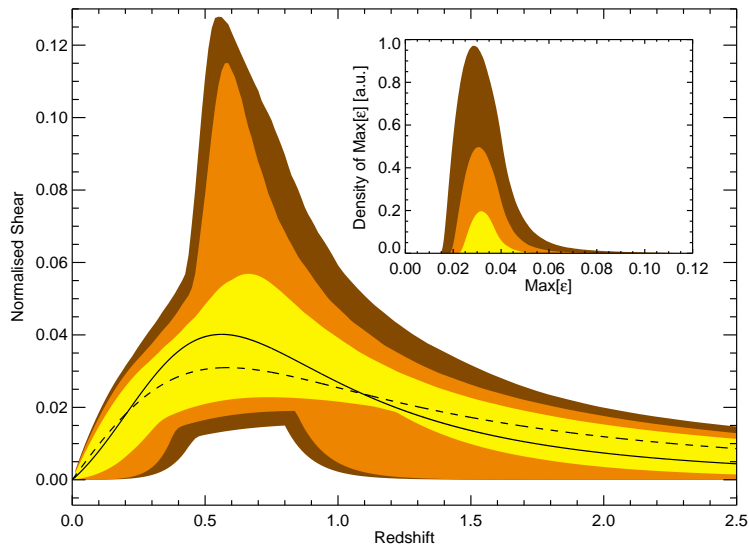}
\caption{The redshift dependence of the normalized shear~(\ref{nshear}) 
for an LTB universe with cosmological parameters corresponding to the best-fit model
(full line) and to the model with minimum $\chi^2$ (dashed line) with 1, 2, and 3-$\sigma$
envelopes, that accommodates current CMB, SNIa
and RBAO observations. We also plot in the inset the distribution function of maximum
shear in 1, 1+2 and 1+2+3-$\sigma$ bins.}\label{fig:shear}
\end{center}
\end{figure}

In Fig.~\ref{fig:shear} we have plotted the normalized shear $\varepsilon$
for a LTB model with cosmological parameters corresponding to the best fit 
model that accommodates current CMB, SNIa and RBAO observations (see 
section~\ref{sec:obs}). 
A survey that can measure 
$\varepsilon$ in Eq.~(\ref{nshearapp}) significantly above zero will be able to distinguish
homogeneous FRW models from inhomogeneous universes. The maximum value of
the shear is attained when the density gradient is maximal, and thus the redshift of the
maximum gives information about the size of the void. Note that the typical value of the 
normalized shear can be small, of order a few percent for LTB models, so this may 
impose severe requirements on future proposed surveys. A way to characterize 
the ability of a survey to distinguish between homogenous and inhomogeneous
models of the Universe would be to ask them to detect a nonvanishing shear
(e.g. ${\rm max}(\varepsilon)\simeq0.03$) at 95\% confidence level, see inset of
Fig.~\ref{fig:shear}. Note also that 
the sign of the (normalized) shear parameter can be used to characterize 
and distinguish a void from an overdensity.

\section{Standard Rulers in LTB models}\label{sec:scales}
Many bounds from observational cosmology, such as the sound horizon (a ``standard
ruler''), the CMB, and applications of the Alcock-Paczinsky test, are derived by
considering and calculating physical scales and processes in the early universe,
and are based on the implicit assumption of an underlying FRW metric. To
test LTB models against these observational data we have to connect distance
scales, redshifts, and expansion rates in the early universe to those observed
today.

Starting from an approximately uniform universe at a high redshift
$z_e$ in the LTB model, the expansion rate and matter density becomes
gradually inhomogeneous, and a uniform comoving physical length scale
$l$ in the early universe at $z_e$, for example the sound horizon, is
not uniform at some later redshift $z$, because of the space dependent
expansion rate. In particular the comoving length at the current time $t_0$
depends on how much relative expansion there
has been at different positions since the formation of the uniform length scale
\begin{equation}
l(r(z)) = l(r_\infty) \frac{A(r(z),t_0)}{A(r(z),t(z_e))}
\frac{A(r_\infty,t(z_e))}{A(r_\infty,t_0)}\,,
\end{equation}
where $r_\infty$ is the radial coordinate of an observer far away
from the void. While it is clear that the {\em physical} length becomes
scale dependent due to the inhomogeneous expansion of the Universe, 
the scale-dependence of the {\em comoving} scale is a consequence of it 
being measured at $t_0$. If instead we {\em defined} the comoving
length scales to be measured in the early universe at $t(z_e)$, then
indeed $l(r(z))$ would be independent of the observer position.
The convenience of the above formula is, that the LTB models we consider
in this paper are asymptotically Einstein de Sitter, and we can easily compute
comoving scales at infinity.

Not only do we have to take into account that scales are different, but we
would also like to use the standard framework to compute early universe
quantities. Hence, we need to compute the equivalent Hubble constant
at the current redshift, that an observer in a Einstein de Sitter universe would
have now at $t_0$ to have the same redshift history far away from the void, that the
central observer has in the LTB model. Using the normal FRW equation for
an Einstein de Sitter universe we can write:
\begin{equation}\label{eq:heff}
H^2(z_e) = H_{\rm eff}^2 (1 + z_e)^3 \,
\end{equation}
where $z_e$ is a high enough redshift where the lightcone is far away from 
the void, but where the radiation density has not become significant yet,
typically $z\simeq100$. The l.h.s.~has to be evaluated
in the LTB model, while the r.h.s.~gives the equivalent
Einstein de Sitter Hubble constant $H_{\rm eff}$, that can be used as input
for calculating e.g.~the sound horizon using standard methods\footnote{See
\cite{GBH:2008a} for the general case when the space time is not asymptotically
Einstein de Sitter, and Ref.~\cite{Zibin:2008b} for a treatment, that also fixes the
angular diameter distance to the correct value in the effective Einstein de Sitter
model, so the full angular powerspectrum can be computed by standard CMB packages}.

\section{Constraints from observations}\label{sec:obs}
To constrain the parameters of our model we use three data sets: The first acoustic peak
in the CMB as measured by WMAP~\cite{WMAP3}, the Union Supernovae data
set~\cite{UNION} consisting of 307 Type Ia supernovae, and the radial baryon acoustic
scale measured at $z=0.24$ and $z=0.43$ using large red galaxies in the SDSS
DR6~\cite{RBAO,GMS}.

The likelihood for the acoustic CMB scale is calculated as in \cite{GBH:2008a}. For
the UNION supernovae we have used the full covariance matrix published by the
Supernova Cosmology Project \cite{UNION,SCP} including systematic errors, and
we have adapted their likelihood code to our {\tt easyLTB} program. We find the radial
baryon acoustic differential redshift $\Delta z$ in our model as
\begin{equation}
\Delta z_{LTB} = \frac{H_L(z) r_s(z)}{c}
\end{equation}
and compare it to the measurements, including systematic errors, published
in \cite{GMS}. The size of the sound horizon is redshift dependent and computed
as detailed in section \ref{sec:scales}.

\begin{table}[h]
\begin{center}
\begin{tabular}{@{}c|c@{ }c@{ }c|c|c@{ }c}
\hline \hline
      & $H_0$ & $H_{r=0}$ & $H_{r=\infty}$ & $\Omega_{\rm in}$
      & $r_0$ & $\Delta r$ \cr
{\footnotesize units}
      & \multicolumn{3}{c|}{{\footnotesize 100 km s$^{-1}$ Mpc$^{-1}$}} &
      & {\footnotesize Gpc} & $r_0$ \cr
\hline
Priors                   & $0.50\!-\!0.95$ & $0.4\!-\!0.89$ & $0.33\!-\!0.63$ & $0.05\!-\!0.35$
                         & $0.5\!-\!4.5$ & $0.1\!-\!0.9$ \cr
Best Fit $\pm$ 2-$\sigma$   & $0.67\!\pm\!0.03$ & 0.58 & 0.45 & $0.16\!\pm\!0.09$
                         & $2.7\!\pm\!0.8$ & $0.44 (>\!0.12)$ \cr
Minimum $\chi^2$ & $0.68$ & 0.59 & 0.45 & $0.145$
                         & $2.35$ & $0.85$ \cr
\hline \hline
\end{tabular}
\end{center}
\caption{Priors used when scanning the parameters of the model, the marginalised best
fit values with 2-$\sigma$ confidence levels (see Fig.~\ref{fig:likelihood1d}), and the
values at the minimum $\chi^2$. $H_0$ is only a pre factor for
$H_0(r)$ and the priors on $H_{r=0}$ and $H_{r=\infty}$ are derived from the
priors on $\Omega_{\rm in}$ and $H_0$. Because they have a complicated prior,
we give them without confidence limits.}\label{tab:priors}
\end{table}

\begin{figure}
\begin{center}
\includegraphics[width=0.65 \textwidth]{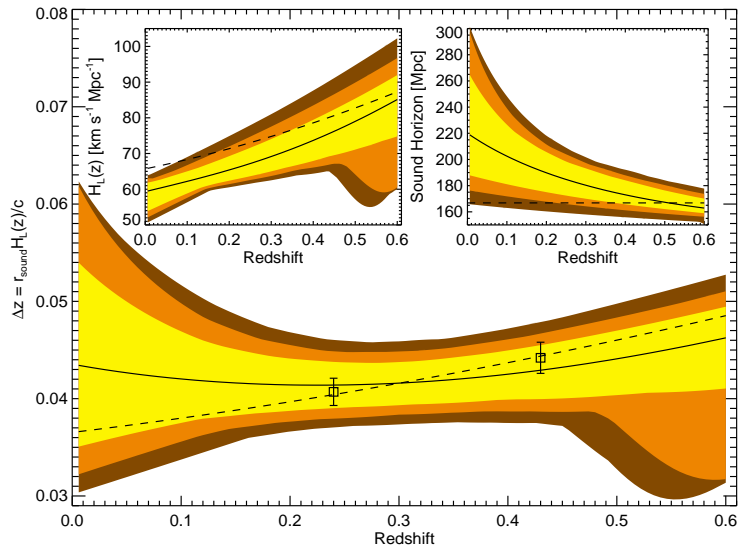}
\caption{The 1, 2, and 3-$\sigma$ limits for the sound horizon differential
redshift $\Delta z$ along the line of sight, the longitudinal Hubble rate $H_L$,
and the sound horizon together with the best fit model (full line). The observed
$\Delta z$ with errors are given as boxes, and the equivalent best fit
$\Lambda$CDM model from \cite{RBAO,GMS} is indicated by a dashed line.
}\label{fig:bands}
\end{center}
\end{figure}

\begin{figure}
\begin{center}
\includegraphics[width=0.9 \textwidth]{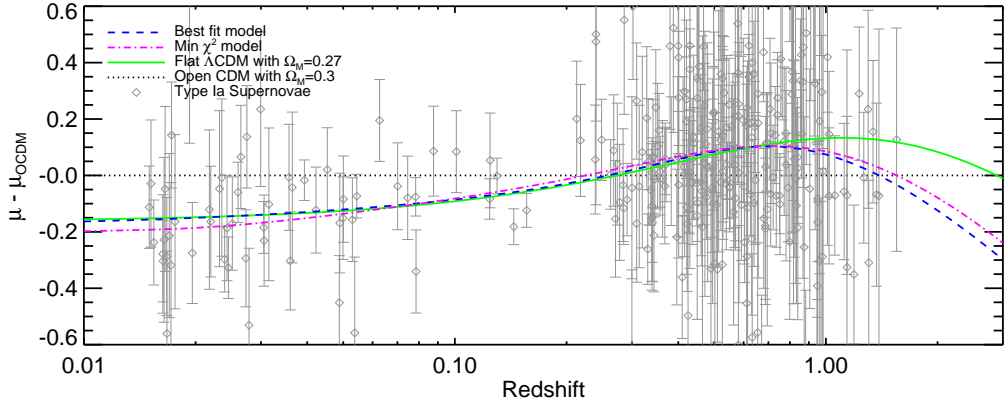}
\caption{The UNION supernova data together with the magnitude-redshift relations for
the best fit and minimum $\chi^2$ LTB models compared to the standard
$\Lambda$CDM model.
}\label{fig:sne}
\end{center}
\end{figure}

\begin{figure}
\begin{center}
\includegraphics[width=0.48 \textwidth]{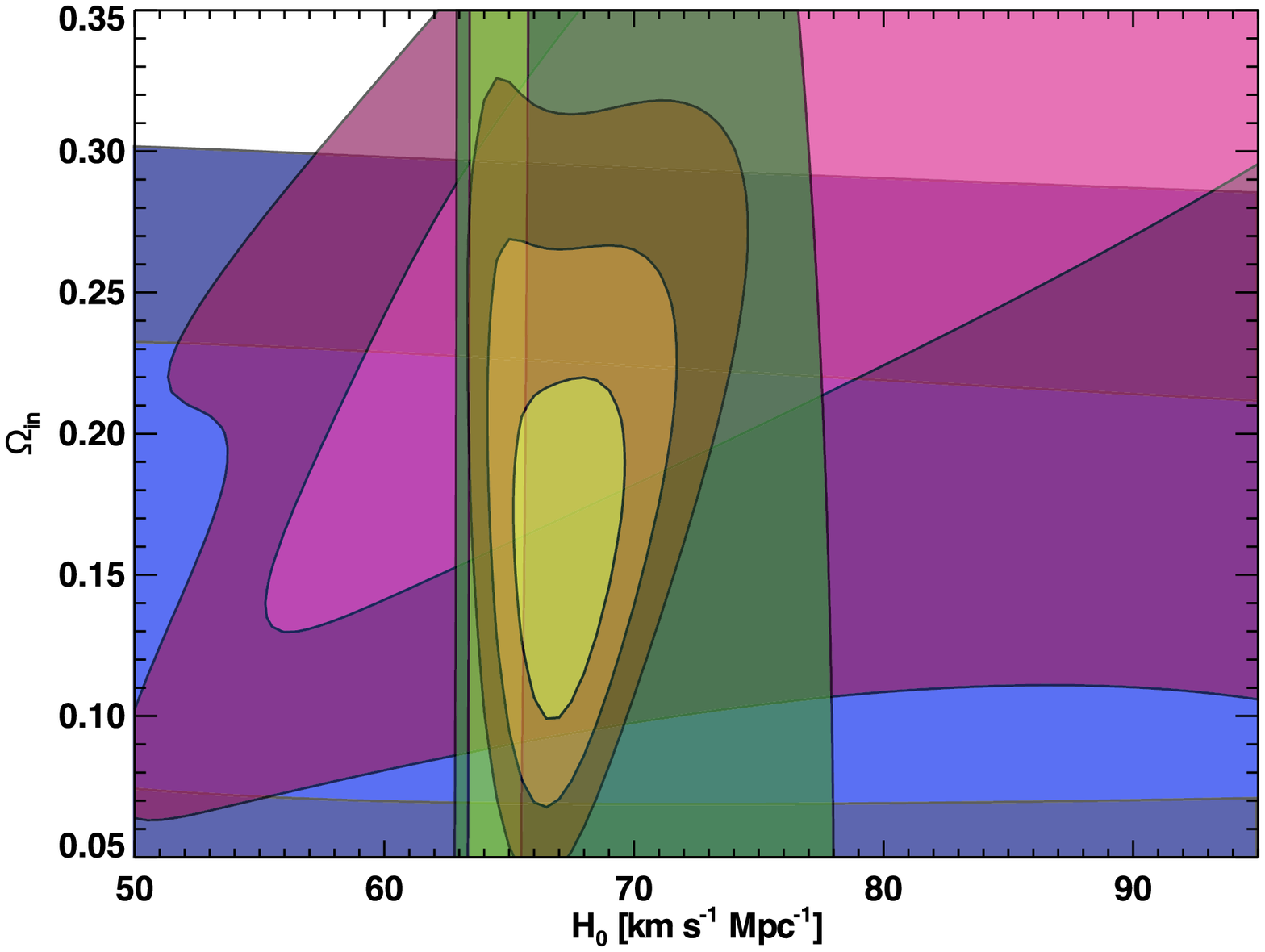}
\includegraphics[width=0.48 \textwidth]{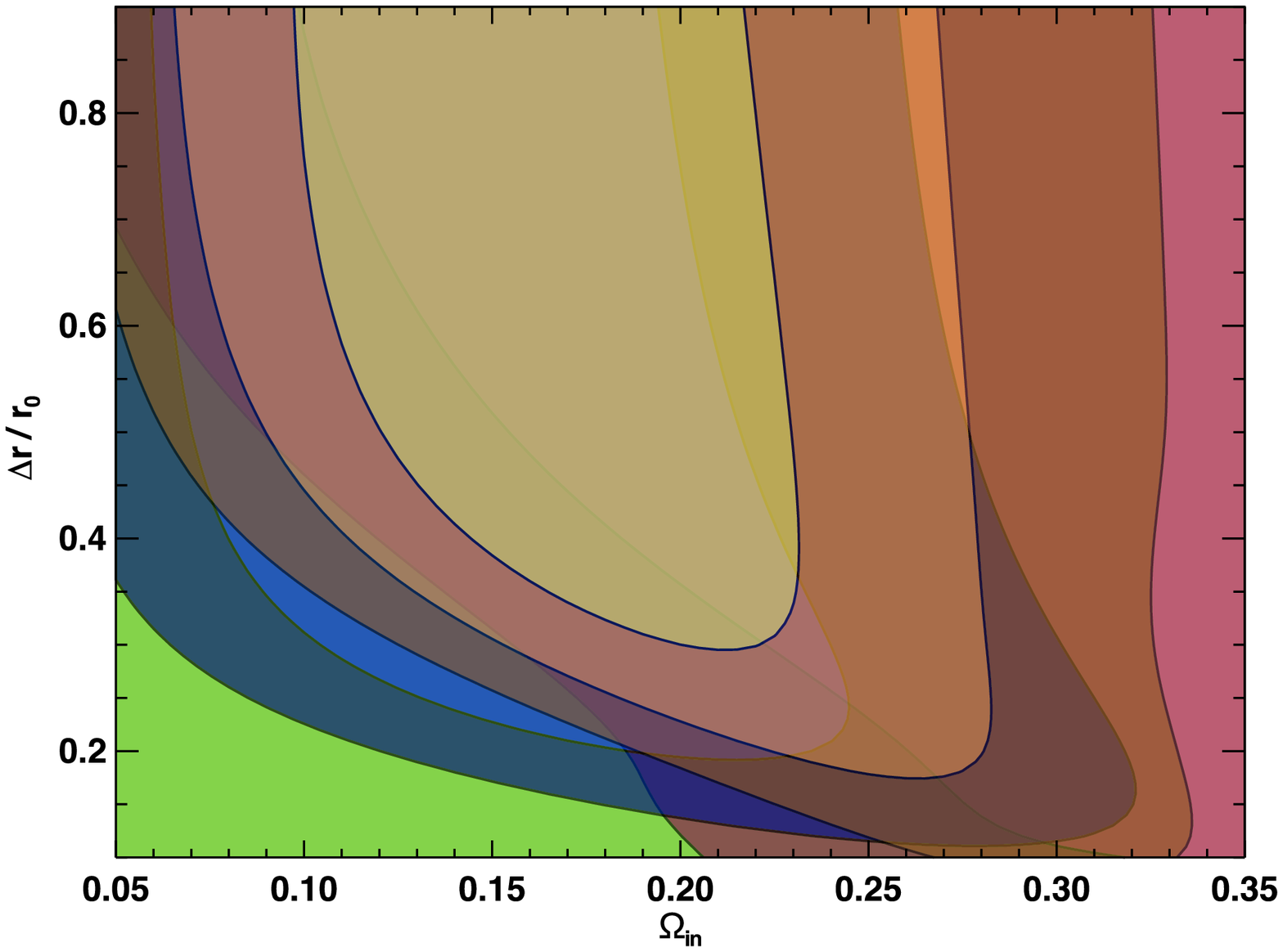}
\includegraphics[width=0.48 \textwidth]{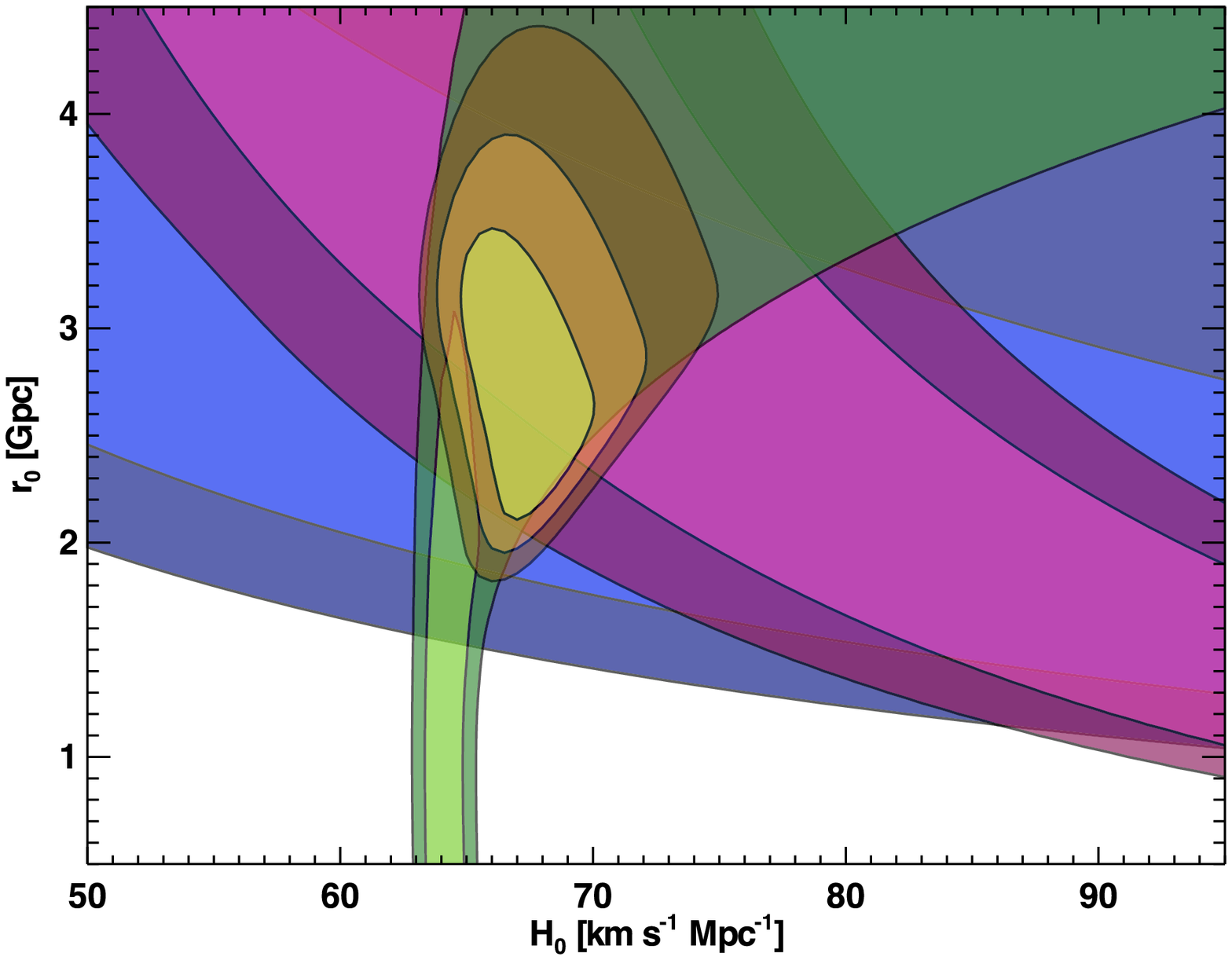}
\includegraphics[width=0.48 \textwidth]{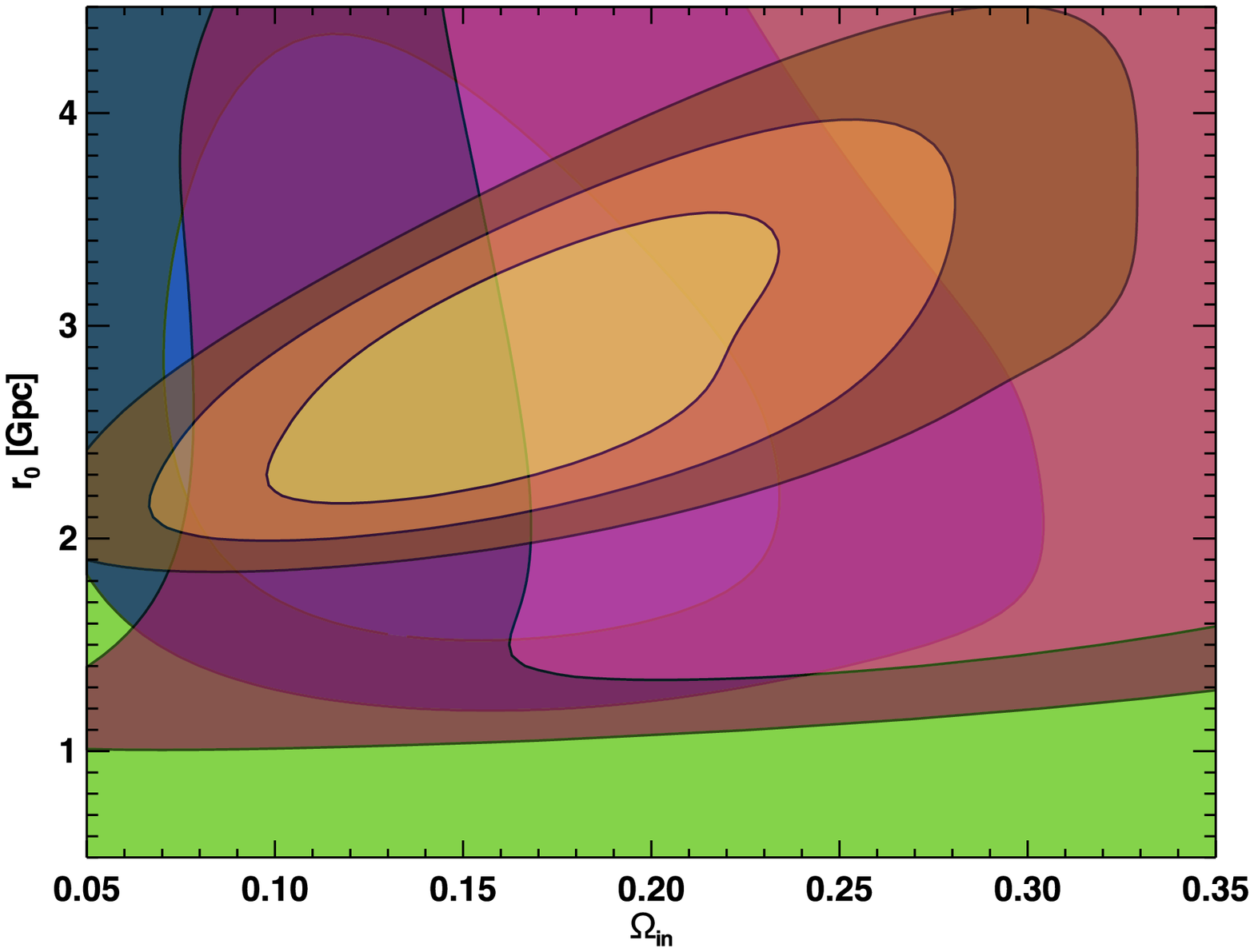}
\includegraphics[width=0.48 \textwidth]{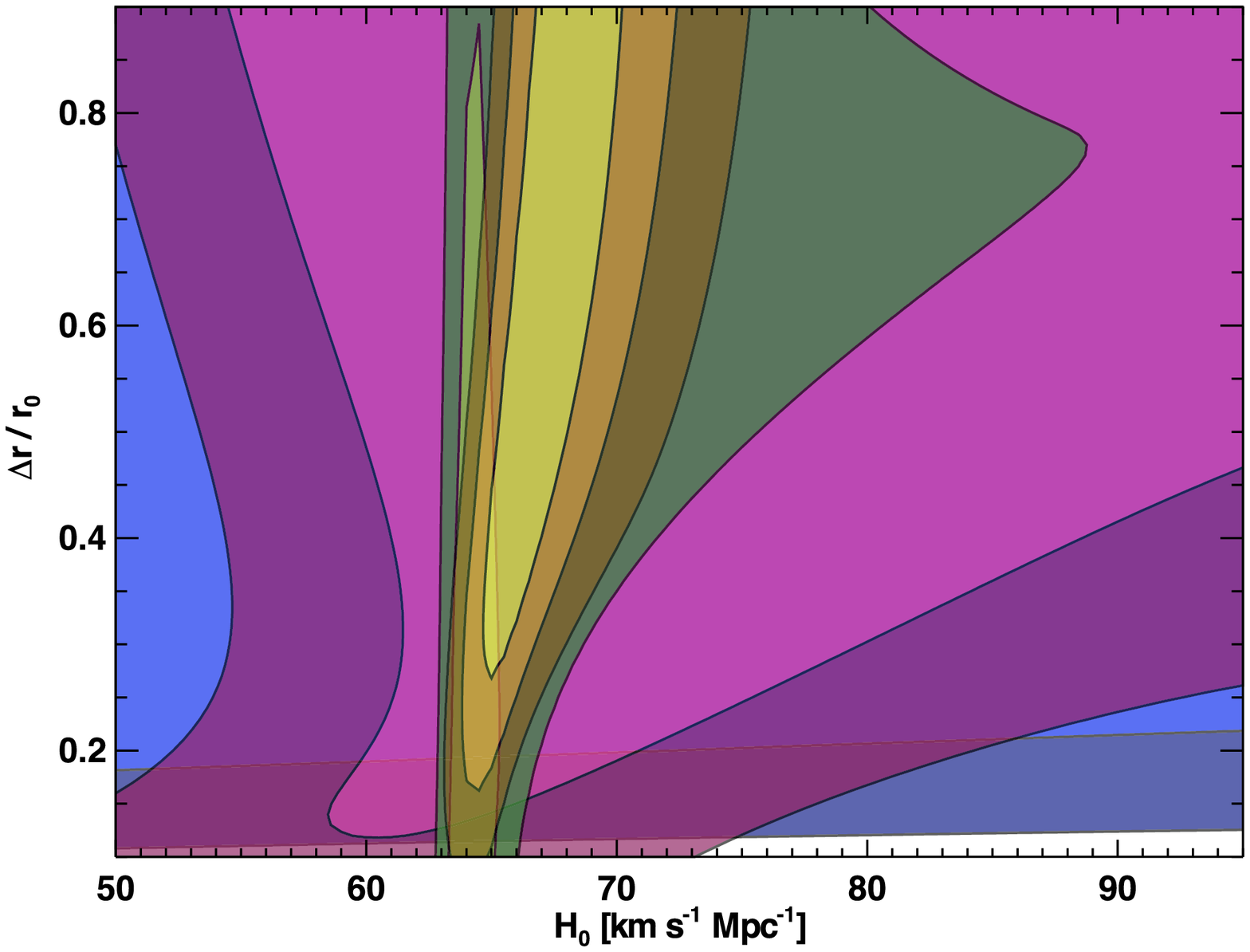}
\includegraphics[width=0.48 \textwidth]{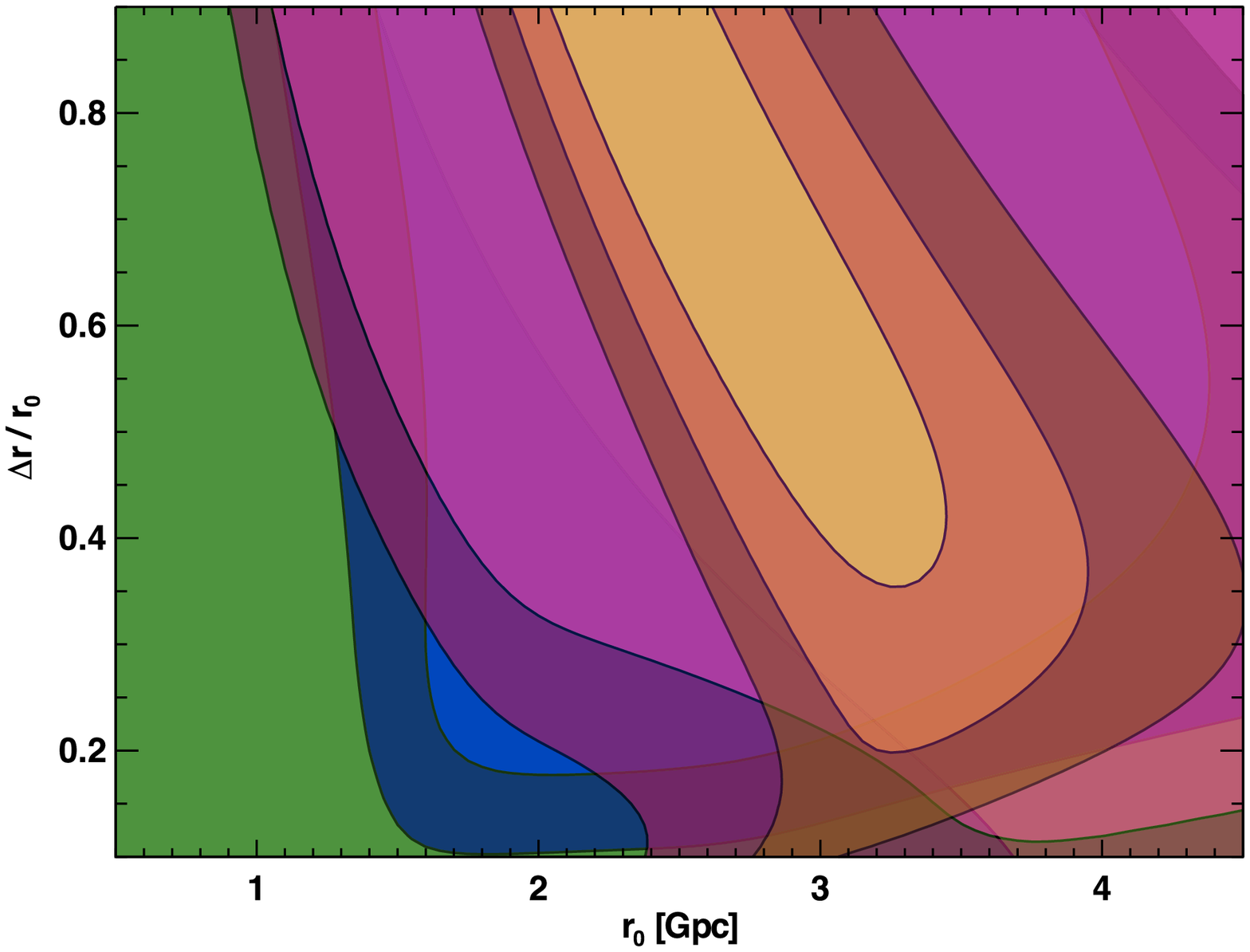}
\caption{Likelihood contours for the combined data set is
shown in yellow with 1-, 2-, and 3-$\sigma$ contours, while the
individual likelihoods for the SNIa, BAO, and CMB data sets are
shown in blue, purple, and green respectively with 1- and
2-$\sigma$ contours.}
\label{fig:likelihood2d}
\end{center}
\end{figure}

Several changes can be observed (see Fig.~\ref{fig:likelihood2d}) compared to our previous
analysis in \cite{GBH:2008a}: Clearly the UNION supernovae prefer a higher overall density
contrast of the void, compared to the Davis et al. compilation \cite{Davis:2007na}. This is
curious, given that most of the supernovae are overlapping between the two data sets, and it
is not a priori clear where the difference is. A possible cause, apart from the increased number
of supernovae in the UNION set, could be that all supernovae in the UNION compilation has
been reanalysed with the SALT light curve fitter, presumably giving a more homogeneous data
set without spurious offsets between different groups of supernovae. The magnitude-redshift
relation is shown in Fig.~\ref{fig:sne}, and even though we only imposed as good a fit as possible
to the data, the curves of the LTB models with the best fit, and the minimum $\chi^2$ are both
almost identical to the best fit $\Lambda$CDM model, illuminating the very small freedom
in the magnitude-redshift relation allowed by current data. Notice also that at $z>2$ the
slopes of the three curves are approximately the same, because at those redshifts all models are
effectively Einstein de Sitter. Hence, a $\sim$5 Gpc sized void would be able to mimic the
$\Lambda$CDM magnitude-redshift relation up to any redshift.

The radial baryon acoustic (RBAO) scale is pushing the void size upwards compared to
the isotropised BAO data of Percival et al \cite{Percival:2007yw}. This can be understood in
terms of the higher redshift of the second data point (z=0.45), compared to the second data
point of Percival et al (z=0.35). Given that the RBAO measurements are in perfect agreement
with a $\Lambda$CDM model, the void has to be bigger, to have the same RBAO differential
redshifts, $\Delta z$, at the two different redshifts. In Fig.~\ref{fig:bands} is shown the envelope curves of
$\Delta z$, $r_s(z)$, and $H_L(z)$ as a function of redshift derived from models which are
inside the 1, 2, and 3-$\sigma$ likelihood contours. While the best-fit LTB model is perfectly
inside the error bars of the two data points, it is very different from the $\Lambda$CDM model
at higher and lower redshifts, and just another redshift bin, outside current measurements,
would help considerably in discriminating between the models. While $\Delta z$ is monotonically increasing
in $\Lambda$CDM, it is almost constant at $z<0.5$ in the best-fit LTB model. This can be
understood by looking at the inserts in Fig.~\ref{fig:bands}: The sound horizon is much larger at
lower redshift, because of the higher expansion rate near the centre of the void, and this is 
almost compensated by the longitudinal Hubble rate, giving a practically flat RBAO differential 
redshift. Unfortunately it is difficult due to measure the RBAO feature at lower redshifts, because
of cosmic variance, but in the near future dedicated RBAO surveys will detect it at higher redshifts.

\begin{figure}
\begin{center}
\includegraphics[width=0.48 \textwidth]{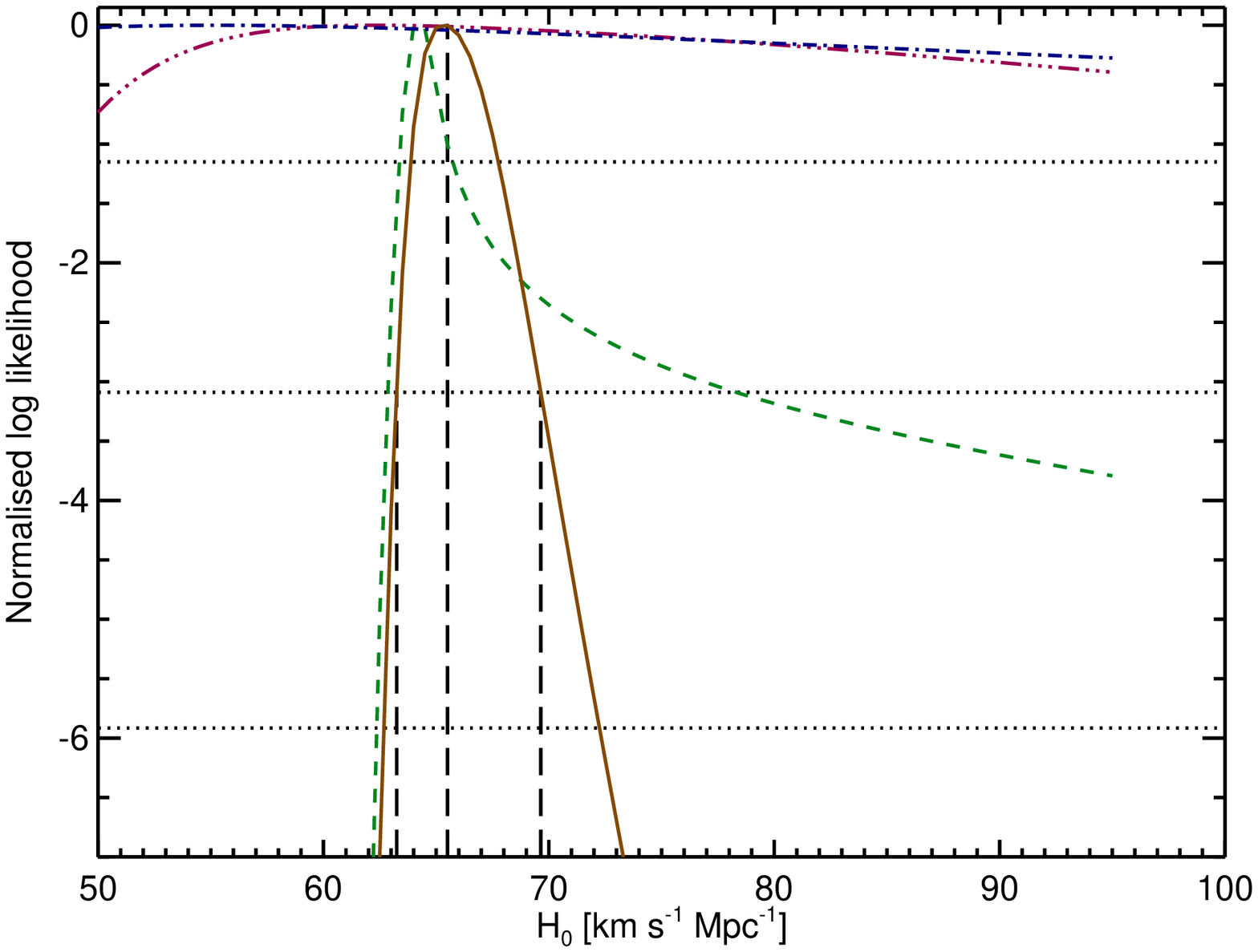}
\includegraphics[width=0.48 \textwidth]{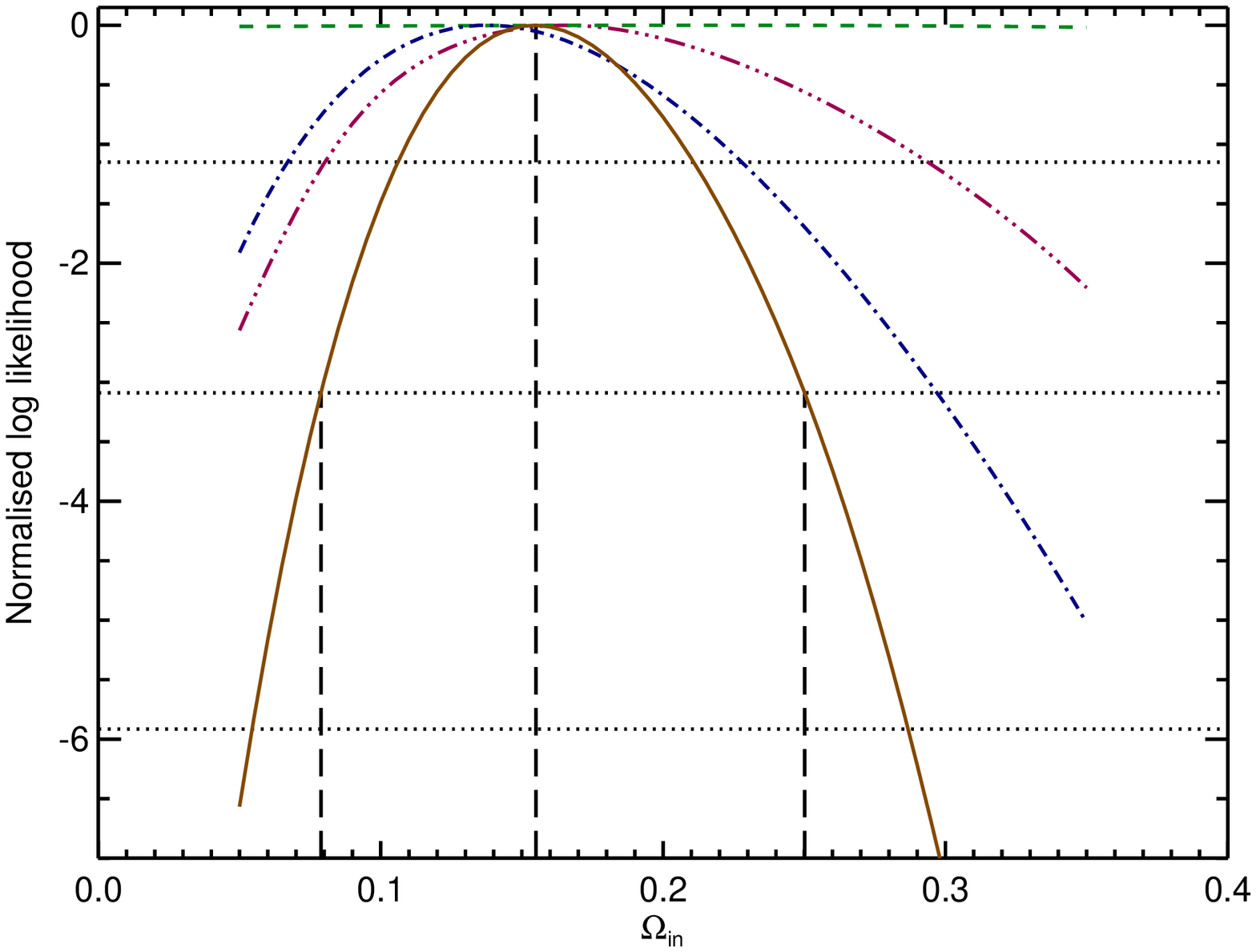}
\includegraphics[width=0.48 \textwidth]{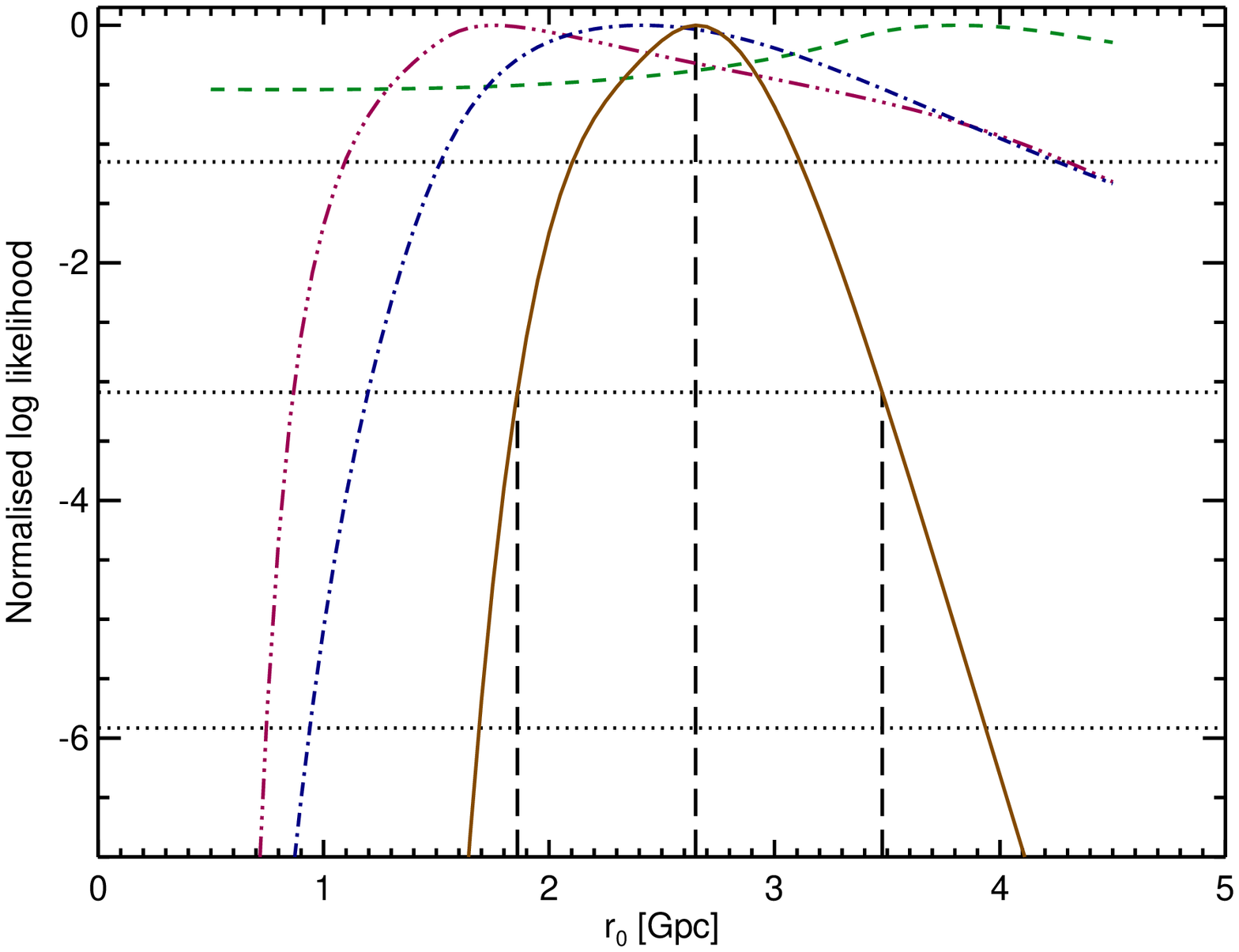}
\includegraphics[width=0.48 \textwidth]{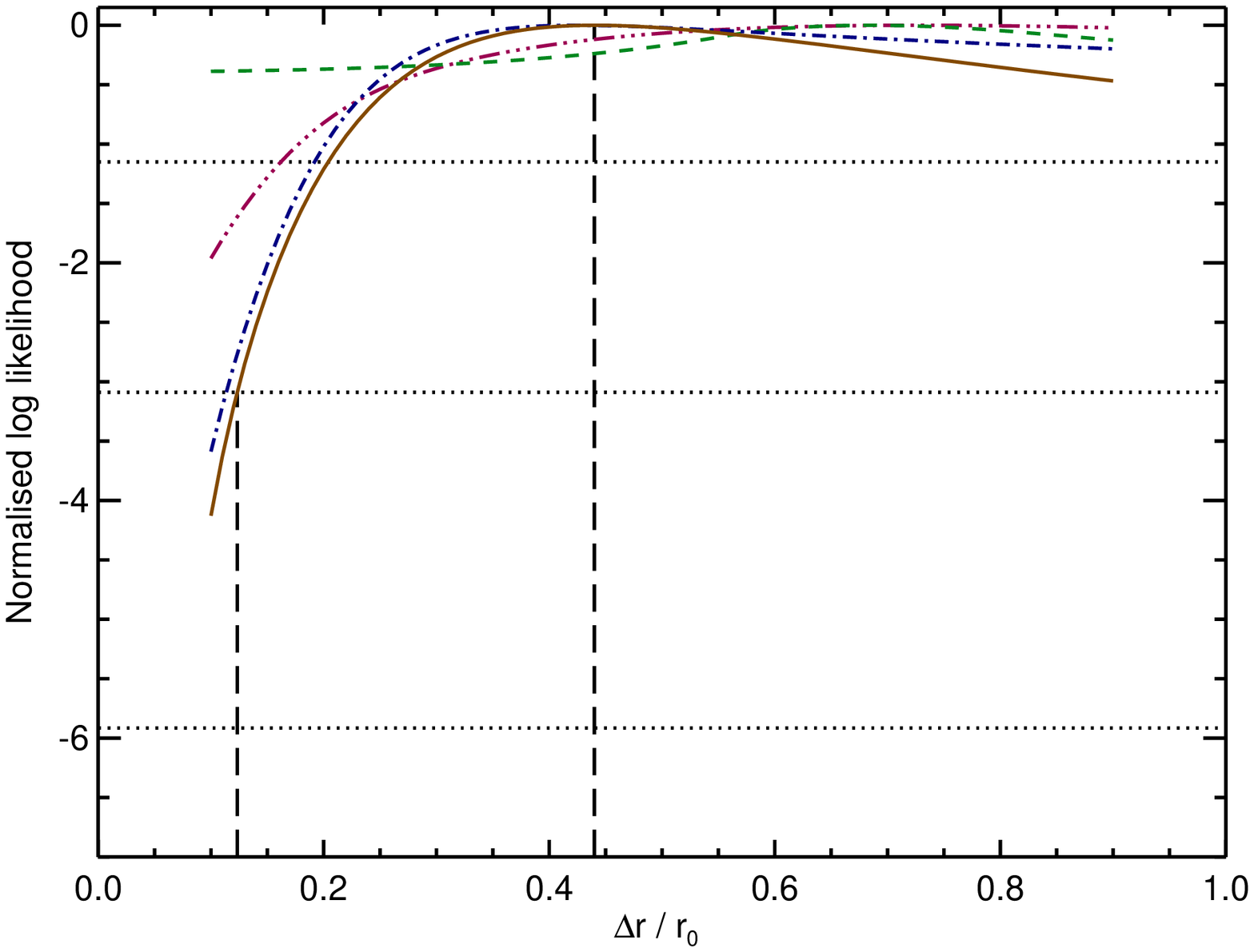}
\caption{Marginalised one dimensional likelihood curves for the combined data set is shown in
yellow (full line), while the individual likelihoods for the SNIa, BAO, and
CMB data sets are shown in blue (dot-dashed), purple (dot-dot-dashed), and
green (dashed) respectively. The dotted lines indicate the 1-, 2-,
and 3-$\sigma$ levels, while the long dashed lines gives the best fit
and the 2-$\sigma$ values.
}
\label{fig:likelihood1d}
\end{center}
\end{figure}

The CMB acoustic scale fixes the overall Hubble parameter $H_0$, because our models are asymptotically
flat, and because we have fixed $\omega_{\rm baryon} = 0.0223$ to the best fit WMAP3
value. Overall the model is capable of yielding a very good fit to data with a minimum $\chi^2 = 307.7$ for 310-4
d.o.f. In figure \ref{fig:likelihood1d} we show the one dimensional likelihoods, and it can be seen that
the different data sets yield consistent and overlapping marginalised likelihoods, further supporting
that the model is a very good and consistent fit to the observations.

\section{Discussion and conclusions}

As an alternative explanation of the dimming of distant supernovae it has recently been 
advocated that we live in a special place in the Universe near the centre of a large void. 
In this scenario, the universe is no longer homogeneous and isotropic and the apparent 
late time acceleration is actually a consequence of spatial gradients in the metric.

If, at the end, local spatial curvature explains away the need for a cosmological constant we 
would be living through a paradigm shift similar to the one that occurred with the discovery of
the expansion of the universe by Hubble and others. At that time, the metric of the universe 
was assumed to have maximal symmetry (time and space translations plus rotations), 
suggesting a static, homogeneous and isotropic universe, in agreement with the Perfect
Cosmological Principle. The observed expansion of the universe removed the need for time 
translation invariance and only spatial sections were maximally symmetric, and we are left with
a homogeneous and isotropic universe satisfying the usual Cosmological Principle, which
states that any observer in the universe is equivalent to any other one. However, if
the universe around us is actually curved, within a large void in an otherwise EdS universe, 
then it is no longer homogeneous, and we are left only with rotational invariance. The metric 
corresponds to that of a Lemaitre-Tolman-Bondi model. These models can accommodate the
observed dimming of distant supernovae, without the need for a cosmological constant,
because spatial curvature also increases the luminosity distance. In fact, LTB models fit well 
the present SN data, even better than $\Lambda$CDM for some models~\cite{Clifton:2008hv}.

In this paper we have tested a class of simple LTB models, against
different cosmological data, that probes the overall geometry of the universe, as observed in
the lightcone, namely: The Union supernovae data set of 307 Type Ia SNe \cite{UNION}, 
the recent measurement of the radial BAO scale in two different redshift bins \cite{RBAO},
and the CMB acoustic scale as measured by WMAP \cite{WMAP3}. We have shown that
LTB models yield a convincing fit to current observations with only 4 parameters in the model,
and an excellent $\chi^2$ for the best-fit model of $\chi^2 / {\rm d.o.f.} = 307.7/(310-4)$.
While not ruling out the void model, the observations do require a void of at least
$r_0 \geq 2$ Gpc (at 95\% C.L.), and a central underdensity that is in accordance with
local galaxy survey measurements of $\Omega_{\rm in} \approx 0.16-0.25$.
The predicted local Hubble parameter of 59 km s$^{-1}$ Mpc$^{-1}$ is within 2-$\sigma$
of the Hubble key project value of $72\pm8$ km s$^{-1}$ Mpc$^{-1}$.

Moreover, there are new specific predictions that these inhomogeneous models give for 
physical observables, the most 
important one being that the overall rate of expansion is not homogeneous and thus 
longitudinal and transverse expansion rates along the line of sight are different. In the language of
differential geometry, this implies that the congruence of geodesic observers has a 
non-vanishing shear. If we normalize this shear w.r.t. to the expansion (they both appear 
on equal footing in the Raychaudhuri equation that governs the acceleration of the
universe) then the normalized shear can be used as an observable for future surveys.
We have found a very compact form of the normalized shear that can be measured in 
principle, written in terms of directly observable quantities. In fact, if future surveys 
measure at high confidence level a non-vanishing value for this new observable, then 
we will have to accept that we live in an inhomogeneous universe, since any 
homogeneous model (be it $\Lambda$CDM or any other generalized vacuum energy 
models) has vanishing shear. In a particular class of inhomogeneous models describing 
large spherical voids, also known as LTB models, the normalized shear has a maximum 
at the position of the steepest gradient in the void profile. Supernovae observations are
compatible with an LTB 
model with void size of order a few gigaparsecs, well beyond present galaxy catalogs
but within reach of the next generation. In those models, the maximum shear is of order
a few percent, and may be difficult to measure over the whole redshift range. In the future,
the ability of a galaxy survey to measure the normalised shear could be used as a 
discriminator between different proposals, in a way similar to (and independent of) the 
figure of merit for the characterization of the nature of dark energy in terms of an
equation of state parameter and its derivative, as advocated by the Dark Energy Task Force.

Remote measurements of the CMB anisotropies through precise determinations of the
kinematic Sunyaev-Zel'dovich effect in a handful of clusters, as recently proposed by
us \cite{GBH:2008b}, or via spectral distortions in the local CMB radiation induced by
CMB photons passing through the void that are later rescattered back towards us during
reionisation, proposed by Caldwell and  Stebbins~\cite{Caldwell:2007yu} are other
new ways of testing the Copernican Principle and constraining general void models.

It is worth mentioning here the recent independent observations of a systematically large 
bulk flow on large scales both above and below 100
Mpc~\cite{Kashlinsky:2008a,Kashlinsky:2008b,Watkins:2008hf}, 
based on the kinematic Sunyaev-Zel'dovich effect in X-ray clusters, and various peculiar 
velocity surveys in the IRAS-PSCz density field respectively. The size of the flow is
significantly above 
that predicted in the Standard Model and constitutes a challenge for $\Lambda$CDM.
Whether this new observation is due to a large void at distances of order gigaparsecs
is still very uncertain, and we will probably have to wait for the next generation of galaxy
surveys to rule out this possibility.

The combination of the steadily increasing body of Type Ia Supernovae,
constraining the transverse Hubble rate, together with upcoming galaxy
surveys that will measure the radial BAO scale in many redshift slices,
such as PAU and BOSS, constraining the longitudinal Hubble rate, will
put significant pressure on LTB models, or measure for the first time the shear
component of cosmic expansion. If not confirmed by observations, then
only a small corner of parameter space -- with very smooth void profiles
and $H_T\approx H_L$ -- will be left.

{\em Note added}: When we were finalizing this work, we learned of a similar analysis 
made by Zibin et al.~\cite{Zibin:2008b}, where they also used the UNION supernova and
the radial BAO of Ref.~\cite{RBAO,GMS} to constrain void models. They also tested their
void model against the CMB anisotropies, and used a toy model for the ISW effect.
They get similar void sizes, albeit they have bigger difficulties in fitting the data, due to
their inclusion of the anisotropies in the CMB.

\section*{Acknowledgments}
We thank James Zibin, Tirthabir Biswas, and Alesio Notari for interesting discussions. 
We acknowledge the use of the computer resources at the Danish Centre
of Scientific Computing. We also acknowledge financial support from 
the Spanish Research Ministry, under contract FPA2006-05807 and
the Consolider Ingenio 2010 programme under contract CSD2007-00060
``Physics of the Accelerating Universe" (PAU).

\section*{References}
\bibliographystyle{hunsrt}
\bibliography{paper}

\end{document}